# Forecasting time series for power consumption data in different buildings using the fractional Brownian motion

Valeria Bondarenko, Simona Filipova-Petrakieva, Ina Taralova, Desislav Andreev

*Abstract*—In the present paper will be discussed the problem related to the individual household electric power consumption of objects in different areas – industry, farmers, banks, hospitals, theaters, hostels, supermarkets, universities. The main goal of the directed research is to estimate the active $P$ and full $S$ power consumptions for all studied buildings. The defined goal is achieved by solving of the following three problems. The first problem studies which buildings increase their power consumption. The second one finds which objects have the greatest increase of power consumption. And the third problem regards if it is possible to make a short-term forecast, based on the solutions of previous two problems. The present research and solving of the aforementioned problems is conducted using fractional Brownian motion theory. The applicability of this approach is illustrated on the example with 20 real objects in different areas. The paper ends with conclusion notes about possibilities to make short-term forecasts about power consumption of the considered buildings.

*Index Terms*—stochastic model, fractional Brownian motion (fBm), parameters estimation, forecasting power consumption

## I. INTRODUCTION

THE act of predicting the variation of some index characterizing the behavior of a particular object for a future period of time is named forecast. It uses data for the considered index for previous moments, which form the time series. Typically, the forecasts are based upon specific assumptions, such as targeted prospects or a preliminarily defined strategy. There exist different mathematical approaches for forecasting: stochastic, intervals, based on artificial intelligent: cluster analysis, neural networks etc. To make a more adequate forecast, on one side it is necessary to have initial data as huge as possible. On the other side, it makes longer the processing time and it requires more powerful computers.

These forecasts could be very useful for different applications such as optimizing the active and the full power energy distribution as a function of the consumption.

There are two types of forecasts: short-time and long-time ones. It depends on the type of the considered object, respectively the associated forecasting parameter. In particular the reliability of the forecast depends on the type of the considered time series and on the amount of the primary data for the analyzed index.

The paper is organized as follows. In the next section are defined three basic problems for analysis of real data about the consumed active power $P$ and the full power $S$ for 20 different objects. The basis of fractional Brownian motion (fBm) theory is presented in Section 3. In Section 4 is discussed the algorithm for analyzing the considered time series. The real examples for the consumed active power $P$ and the full power $S$ are discussed in Section 5. The paper ends with concluding remarks about power consummation and its future forecasting for the considered objects.

## II. PROBLEM STATEMENT

Let us consider the power consumption (active power $P$ and full power $S$) in different administrative buildings where the consumption has different daily and nightly behavior. The analyzed objects correspond to different life's areas such as: 'Bank', 'Automobile Industry', 'BPO Industry', 'Cement Industry', 'Farmers 1', 'Farmers 2', 'Health Care Resources', 'Textile Industry', 'Poultry Industry', 'Residential (individual)', 'Residential (apartments)', 'Food Industry' ,'Chemical Industry', 'Handlooms', 'Fertilizer Industry', 'Hostel', 'Hospital', 'Supermarket', 'Theatre' and 'University'.

The main aim of the research is to estimate the actual power consumption ($P$ and $S$) of each analyzed object and to make forecast about their future consumptions. This aim is realized by solving the following problems:

Problem 1: Which buildings increase their consumption ($P$ and $S$)?

This work has been carried out as a result of the collaboration between Bulgarian and French teams working together during the joint project "Cluster analysis of aperiodic time series signals" financed by the Bilateral program PHC "RILA" projects 2017-2018 – Bulgaria – France.

V. Bondarenko, Research assistant with in the University of Beira, Covilha, (valeria_bondarenko@yahoo.com).
S. K. Filiova-Petrakieva, Member #94362473, IEEE, Bulgarian Section, (Technical University of Sofia, Bulgaria, department "Theory of Electrical Engineering", Faculty of Automation, Sofia, Bulgaria (petrakievas-te@tu-sofia.bg).
I. Taralova, Laboratoire des Sciences du Numerique de Nantes, Ecole Centrale de Nantes, Nantes, France (ina.taralova@ec-nantes.fr)
D. A. Andreev, Technical University of Sofia, Bulgaria, department "Computer Science", Faculty of Computer systems and Technologies; Senior software engineer in department: "Advance Driver Assistance Systems", company: Visteon Electronics Bulgaria, Sofia, Bulgaria (desislav.andreev@gmail.com).

Problem 2: Which buildings have the greatest increase of their consumption (*P* and *S*)?

Problem 3: If it is possible to make a short-terms forecast, based on the results from the solutions of Problems 1 and 2?

### III. BASIC CONCEPTIONS IN FRACTIONAL BROWNIAN MOTION (fBM) THEORY

Consider an observed trajectory $x(t), t \in [0;T]$, which is describing the stochastic evolution of some dynamic object. The mathematical model of this trajectory is defined as a random process $\xi(t)$, where:

$$x(t) = X(t),$$

where $X(\bullet)$ is relization of the process $\xi$.

As a rule, it is chosen as a model random process with known characteristics. Direct use of this definition requires a broad class of these processes. On the other hand, this class includes Gaussian and Markov processes. Let us introduce another definition of continuous mathematical models for the observed trajectory $x(\bullet) \in C[0; T]$ using nonlinear conversion.

Definition. Mathematical model of observed trajectory $x(t)$ is a pair $(\Phi, \xi)$, where $x(t) = \Phi(X(\bullet))(t)$, $\xi(t)$ is a random process with known characteristics, $\Phi$ is a reversible conversion in $C(0; t)$.

Let's assume $t = t_k, k = 1,..,n$, $t_k = t_1 + (k-1)\frac{T-0}{n}, t_1 = 0$, $x_k = \Phi(X(\bullet))(t_k)$ is a model of the observed time series $\{x_1, x_2, ..., x_n\}$.

Let us call $\xi$ as a basic process of the model.

One of the most popular Markov models of time series is Gaussian random process, and fractional Brownian motion (fBm). The large use of this process is caused by its "convenient" properties, which are described below.

Fractional Brownian motion is defined as a Gaussian random process with characteristics:

$$B_H(t), \quad E\{B_H(t)\} = 0, \quad B_H(0) = 0$$

$$E\{B_H(t)\} \cdot B_H(s) = \frac{1}{2}\left(t^{2H} + s^{2H} - |t-s|^{2H}\right)$$

Note that with $H = 0.5$ we get a standard Wiener process, where $H$ indicates the Hurst exponent.

The smoothness of the trajectories of the process $B_H(t)$ is defined by the parameter H: almost all the trajectories satisfy the Holder condition for regularity:

$$|X(t) - X(s)| \le c|t-s|^\alpha, \quad \alpha < H. \tag{1}$$

which generalizes known Levy's result for the Wiener process [2, 3, 4].

The increments of fBm

$B_H(t_2) - B_H(t_1), B_H(t_4) - B_H(t_3), t_1 < t_2 < t_3 < t_4$ form a Gaussian random vector with a correlation between the coordinates:

$$\frac{1}{2}\left((t_4-t_1)^{2H} + (t_3-t_2)^{2H} - (t_4-t_2)^{2H} - (t_3-t_1)^{2H}\right)$$

For discrete time:

$$\xi_k = B_H\left(\frac{k}{n}\right) - B_H\left(\frac{k-1}{n}\right),$$

we obtain the correlation coefficient:

$$\rho(\xi_j, \xi_k) = \frac{1}{2}(|k-j+1|^{2H} + |k-j-1|^{2H} - 2|k-j|^{2H}), \tag{2}$$

It means that increments form a stationary (in the narrow sense) sequence.

Let us mention some properties of fBm:

1. Changing time scale is equivalent to changing the "amplitude" -of the process:

$$Law(B_H(at)) = Law(a^H B_H(t)).$$

This equality denotes the coincidence of one-dimensional distributions of the processes:

$B_H(at)$ and $a^H B_H(t)$.

This property is called self-similarity and it is useful for the analysis of time series.

2. Let us put in the formulae (2) $j = k + n$. Then the correlation coefficient:

$$\rho(\xi_k, \xi_{k+n}) = \frac{1}{2}\left[(n+1)^{2H} + (n-1)^{2H} - 2n^{2H}\right] \tag{3}$$

When $n \to \infty$ it follows:

$$\rho_n = \rho(\xi_k, \xi_{k+n}) \sim H(2H-1)n^{2H-2} \tag{4}$$

So, the increments memory decrease exhibits a power law; the increments are independent for $H = 0.5$. With $H < 0.5$ the increments form the sequence with short memory, and for $H > 0.5$ a sequence with a long memory. The sign of the correlation coefficient $\rho_n$, defined by formula (4), depends on the value of $H$: $\rho_n < 0, H < 0.5$; $\rho_n > 0, H > 0.5$.

For $H < 0.5$ the sequence $\xi_n$ of increments fBm is called "pink noise" and negativity of the variations indicates fast variability values. The process of fBm, $H < 0.5$ is known as "antipersistent".

For $H > 0.5$ the sequence $y_n$ of increments fBm is called "black noise" and the fBm process is known as "persistent".

The original approach using stochastic and statistical analysis is based on real time series modeling assuming a hypothesis that their increments can be modeled by fBm. If the estimated Hurst exponent is bigger than 0.5 the process is called persistent and a forecast can be made.

### IV. THE SELECTION OF THE CONVERSION OF TIME SERIES

It requires initial analysis of the increments $\{y_1, y_2, ..., y_n\}$ to determine the conversion $\Phi$.

In particular, is it necessary to estimate the one-dimensional distribution of the samples and the correlation. These actions are possible only with a large sample size ( $n > 5000$ ). We propose new empirical method of increments transformation $\{y_1, y_2, ..., y_n\}$ into $\{z_1, z_2, ..., z_n\}$ for a small sample, in order to satisfy required statistical properties (e.g. gaussianty).

Let us consider the increments $y_k = x_{k+1} - x_k$, $k = 1,...,(n-1)$ and construct the statistics (kurtosis)

$$d_n = \frac{\left(\frac{1}{n-1} \cdot \sum_{k=2}^{n} |y_k|\right)^2}{\frac{1}{n-1} \cdot \sum_{k=2}^{n} y_k^2},$$

where $d_n$- is equal to $2/\pi$- or Gaussian model.

If $d_n$ is significantly different from $2/\pi$, it is necessary to replace the time series $\{y_1, y_2, ..., y_{n-1}\}$ by the new sequence $\{z_1, z_2, ..., z_{n-1}\}$.

The general idea of approximation is a one-dimensional functional transformation $g$ of each increment $y_k$, where $g$ is an increasing odd function,

$$z_k = g(y_k).$$

Let us assume

$$\lim_{n \to \infty} \frac{\left(\frac{1}{(n-1)} \cdot \sum_{k=1}^{n-1} |g(y_k)|\right)^2}{\frac{1}{(n-1)} \cdot \sum_{k=1}^{n-1} (g(y_k))^2} = \frac{2}{\pi},$$

where $z_k = g(y_k)$ is assumed as a Gaussian random value. Let us demonstrate proposed algorithm with $g$ as a power function.

Assume

$$z_k = \text{sgn } y_k |y_k|^{\frac{1}{\lambda}} = \Phi^{-1}(y_k), \tag{5}$$

$$y_k = \text{sgn } y_k |z_k|^{\lambda} = \Phi(z_k), \lambda > 0, \tag{5a}$$

then

$$d_n = \frac{\left(\frac{1}{n-1} \cdot \sum_{k=1}^{n-1} |z_k|^{\lambda}\right)^2}{\frac{1}{n-1} \cdot \sum_{k=1}^{n-1} |z_k|^{2\lambda}} \tag{6}$$

$d = \lim_{n \to \infty} d_n$ is equal to the ratio of the corresponding mathematical expectations.

If $\xi \sim \Xi(0, \sigma^2) \Rightarrow E|\xi|^{\alpha} = \frac{2^{\frac{\alpha}{2}}}{\sqrt{\pi}} \cdot \sigma^{\alpha} \cdot \Gamma\left(\frac{\alpha+1}{2}\right)$,

then $d = \frac{1}{\sqrt{\pi}} \cdot \frac{\Gamma^2\left(\frac{\lambda+1}{2}\right)}{\Gamma\left(\lambda+\frac{1}{2}\right)}$,

where the parameter $\lambda$ is defined from the equations (5, 5a). Thus, the proposed approximation leads to the following model of original time series:

$$x_k = \sum_{j=1}^{k} \text{sgn}\left(y_j \cdot |z_j|^{\lambda}\right).$$

If we assume that the values of the sequence $\{z_1, z_2, ..., z_n\}$ are increments of fBm, the Hurst exponent can be calculated by the following algorithm, as it can be seen from the proposed method:

1) Construct the statistic:

$$R_1 = \frac{1}{n} \sum_{k=1}^{n} |z_k| = |\bar{z}|$$

2) Calculate the matrix $S_H^{-1}$, where $S_H$ is a correlation matrix of increments fBm:

$$s_{jk} = E\left[B_H\left(\frac{k}{n}\right) - B_H\left(\frac{k-1}{n}\right)\right] \cdot \left[B_H\left(\frac{j}{n}\right) - B_H\left(\frac{j-1}{n}\right)\right] = \rho(\xi_j, \xi_k)$$

3) $$Q = \frac{0.8}{R_1} \sqrt{\frac{\left(S_H^{-1} z, z\right)}{n-1}},$$

The statistics $Q$ **is** calculated for different Hurst exponent values with step (0.05 – 0.1) and the corresponding $\hat{H}$ is calculated, such as:

$$|Q(H)-1| \to \min, \hat{H} = \arg\min |Q(H)-1|. \tag{6}$$

4) Verify the following hypothesis *T*: *The statistics* $\{z_1, z_2, ..., z_n\}$ *which have been obtained by transformation (5) of real data can be simulated by fractional Brownian motion increments.* The algorithm with known *H* is the following. Denote

$$c = \frac{1}{n} \sum_{k=1}^{n} z_k^2,$$

and assume that hypothesis *T* is verified

$$z_k = \sqrt{c}.\xi_k = \sqrt{c}.n^H \left( B\left(\frac{k+1}{n}\right) - B\left(\frac{k}{n}\right) \right).$$

Assume $v_k = \sum_{j=1}^{k-1} z_j$ and construct the statistics [5, 6]

$$A_n = \frac{1}{n} \sum_{k=1}^{n} v_k . z_k^3, \qquad H \in (0; \ 0.5)$$

$$B_n = \frac{1}{n^{1+H}} \sum_{k=1}^{n} v_k^2 . z_k^3, \qquad H \in (0; \ 0.5)$$

$$D_n = \frac{1}{n^{2H}} \sum_{k=1}^{n} v_k . z_k^3, \qquad H \in (0.5; \ 1)$$

If hypothesis $T$ is true, there is convergence [1]:

$$A_n \to -\frac{3}{2}c^2; \quad B_n \to 3c^{2.5}\eta; \quad D_n \to \frac{3}{2}c^2 B^2 \ .$$

The decision about the veracity of the hypothesis $T$ is accepted by comparing the real values of the statistics with their theoretical limit values. Therefore, let us determine the deviation from the limit values for the statistic $A_n$

$$\delta = \left|\frac{A_n - A}{A}\right|.$$

The limit distribution functions for the statistics $B_n$, $D_n$:

$$F_1(x) = P\{3 c^{2.5}\eta < x\} = \Phi\left(\frac{x}{3\sigma}c^{-2.5}\right);$$

$$F_2(x) = 2\Phi\left(\frac{1}{c}\sqrt{\frac{2}{3}}x\right) - 1, \ x > 0,$$

where $\Phi$ is the Laplace function, $\sigma = (2H+2)^{-0.5}$.

The hypothesis $T$ is accepted, if:

$\delta < \beta_0, |B_n| < \beta_1, \ H < 0.5$

or

$0 < D_n < \beta_2, \ H > 0.5$

where $\beta_1, \beta_2$ are quantile distributions from $F_1, F_2$ which correspond to the selected significance level $\alpha = 0.1$.

$$\beta_1 = \frac{4,95c^{2,5}}{\sqrt{2H+2}}, \ \beta_2 = 4,08c^2, \ \text{where } c = \bar{z}^2 \ .$$

## V. REAL EXAMPLES

Let us consider the consumed active power $P$ and full power $S$ for the following 20 objects. The data for individual household electric power consumption are saved each hour [7]. The considered buildings are: 'Bank', 'Automobile Industry', 'BPO Industry', 'Cement Industry', 'Farmers 1', 'Farmers 2', 'Health Care Resources', 'Textile Industry', 'Poultry Industry', 'Residential (individual)', 'Residential (apartments)', 'Food Industry' ,'Chemical Industry', 'Handlooms', 'Fertilizer Industry', 'Hostel', 'Hospital', 'Supermarket', 'Theatre' and 'University'.

The research aim is to estimate the power consumption ($P$ and $S$) of each analyzed object. This aim is realized by solving the following problems:

Problem 1: Which buildings increase their consumption ($P$ and $S$)?

Problem 2: Which buildings have the greatest increase in their consumption ($P$ and $S$)? (7)

Problem 3: If it is possible to make a short forecasting, based on the results from the solutions of Problems 1 and 2?

Before starting the data processing all primary data have been normalized and the respective values of $P$ and S take values in interval [0, 1]. Next, the primary data have been processed using linear approximation of the trend of the initial data and new sequences have been obtained in every example.

This is realized by the following steps:
1. Calculation of the increments

$y_k = x_{k+1} - x_k, \ k = 1,...,(n-1).$

2. Construction of the new sequence

$\{z_1, \ z_2, \ ..., \ z_{n-1}\}$ by (5) and (5a).

3. Hurst exponent estimation by (6).
4. Checking the quality of the approximation by $A_n$ and $B_n$ (7).

The processing results are shown in Tables IA and IB.

As it can be seen from the TABLE IA, for all the data we have antipersistent parameter $\hat{H} < 0.5$. For the examples (except Theater and Food Industry), the condition is clearly verified: $|B_n| < \beta_1$.

For all examples (except Theater and Textile Industry), the condition is clearly verified: $|B_n| < \beta_1$ and the process has short-memory. (8)

According to the results gotten in the previous section it can make the following conclusions about three problems defined in the beginning of the paper:

*Problem 1*: Which buildings increase their consumption ($P$ and $S$)?

The approximation by fractional Brownian motion of the considered real data shows that  Bank, Automobile Industry, BPO Industry,  Textile Industry,  Residential (individual), Chemical Industry, Hospital and University increase their consumption of the active power P, because for these buildings we've got the highest antipersistent Hurst parameter $\hat{H} = 0.4$ (which means that the processing data change more smoothly).

TABLE Iₐ
ACCESSING RATES OF THE CONSIDERED PARAMETERS FOR CONSUMED ACTIVE POWER $P$

| Analyzed object | $\hat{H}$ | $A_n$ | $B_n$ | $A$ | $\beta_1$ | $\beta_2$ |
|---|---|---|---|---|---|---|
| Bank | 0.4 | -1.58 | 0.22 | -1.57 | 2.98 | |
| University | 0.4 | -1.49 | 0.55 | -1.62 | 2.93 | |
| Cement Industry | 0.1 | -1.41 | 1.61 | -1.51 | 3.34 | — |
| Automobile Industry | 0.4 | -1.54 | 0.04 | -1.5 | 2.99 | |
| BPO Industry | 0.4 | -1.72 | -1.61 | -1.51 | 5.28 | — |
| Farmers 1 | 0,3 | -1.42 | 1.68 | -1.5 | 2.59 | |
| Farmers 2 | 0.1 | -3.44 | -6.38 | -3.57 | 9.883 | — |
| Health resources | 0.1 | -1.5 | -1.24 | -1.5 | 3.53 | |
| Textile Industry | 0.4 | 398.8 | -2.125 | -1.5 | 2.95 | — |
| Poultry Industry | 0.1 | -1.47 | -1.58 | -1.5 | 3.82 | |
| Residential (individual) | 0.4 | -1.72 | -1.61 | -1.51 | 5.28 | — |
| Residential (Apartments) | 0.1 | -1.57 | -2.25 | -1.5 | 3.11 | |
| Food Industry | 0,1 | -1.47 | -2.98 | -1.5 | 2.83 | — |
| Chemical Industry | 0.4 | -1.6 | 0.09 | -1.5 | 2.35 | |
| Handlooms | 0.4 | -1.78 | 0.46 | -1.5 | 2.99 | — |
| Fertilizer Industry | 0.1 | -1.56 | -0.28 | -1.5 | 2.87 | |
| Hostel | 0.1 | -1.54 | -0.97 | -1.5 | 2.81 | — |
| Hospital | 0.4 | -1.47 | 0.16 | -1.5 | 2.91 | |
| Supermarket | 0.25 | -1.51 | 0.64 | -1.5 | 2.16 | — |
| Theater | 0.1 | -1.59 | 3.07 | -1.5 | 2.05 | — |

TABLE Iᵦ
ACCESSING RATES OF THE CONSIDERED PARAMETERS FOR CONSUMED FULL POWER $S$

| Analyzed object | $\hat{H}$ | $A_n$ | $B_n$ | $A$ | $\beta_1$ | $\beta_2$ |
|---|---|---|---|---|---|---|
| Bank | 0.4 | -1.34 | 0.31 | -1.47 | 2.77 | |
| University | 0.36 | -1.25 | 0.88 | -1.52 | 2.10 | |
| Cement Industry | 0.24 | -1.38 | 1.92 | -1,29 | 3.01 | — |
| Automobile Industry | 0.39 | -1.59 | 0.2 | -1.68 | 2.57 | |
| BPO Industry | 0.46 | -1,87 | -1,23 | -1.08 | 4,01 | — |
| Farmers 1 | 0.28 | -1.80 | 1.32 | -1.6 | 2.0006 | |
| Farmers 2 | 0,17 | -3.34 | -5.36 | -3.03 | 10.01 | — |
| Health resources | 0.28 | -1.34 | -0.99 | -1.45 | 2.44 | |
| Textile Industry | 0.46 | 201.2 | -118 | -1.8 | 4.5 | — |
| Poultry Industry | 0.18 | -1.59 | -2.21 | -1.48 | 5.77 | |
| Residential (individual) | 0.39 | -1.22 | -1.85 | -1.50 | 6.54 | — |
| Residential (Apartments) | 0.2 | -1.33 | -3.77 | -1.2 | 5.5 | |
| Food Industry | 0.17 | -1.99 | -2.42 | -1.66 | 4.5 | — |
| Chemical Industry | 0.36 | -1.21 | 0.5 | -1.01 | 4.66 | |
| Handlooms | 0.38 | -1.56 | 0.43 | -1.39 | 2.2 | — |
| Fertilizer Industry | 0.21 | -1.88 | -0.44 | -1.5 | 2.56 | |
| Hostel | 0.11 | -1.2 | -0.5 | -1.9 | 2.4 | — |
| Hospital | 0.29 | -1.73 | 0.12 | -1.5 | 2.37 | |
| Supermarket | 0.29 | -1.58 | 0.23 | -1.5 | 2.89 | — |
| Theater | 0.11 | -1.43 | 3.55 | -1.5 | 2.14 | — |

*Problem 2*: Which buildings have the greatest increase in their consumption (*P* and *S*)?

The approximation by fractional Brownian motion of the considered real data shows that Bank, BPO Industry and Textile Industry increase their full power *S* consumption because for these buildings we've got the highest antipersistent Hurst parameters: $\hat{H}=0.4$, $\hat{H}=0.46$ and $\hat{H}=0.46$, respectively (which means that the processing data change more smoothly).

*Problem 3*: Is it possible to make a short forecasting, based on the results from the solutions of *Problems 1* and *2*?

All of the approximations have antipersistent character ($\hat{H}<0.5$) and it is adequate, if the conditions (5) are satisfied. Therefore, it is impossible to construct a short-term forecast. Then other methods will be used for approximation of the analyzed real data to make more reliable prognoses. But it will be the focus of the future investigations, where some newer techniques such as cluster analysis will be applied.

Numerical experiment has shown that for a lot of data it is possible to implement approximation by fractional Brownian motion, but without forecast and with short memory.

For the examples Food Industry, Textile Industry and Theater conditions (5) are not satisfied, so it means that they can not be assumed as Gaussian processes and it is possible to approximate them by other distributions.

## VI. CONCLUSION

Fractional Brownian motion is used for approximation of real object data to estimate a capability of short and long-term prognosis about an active power *P* and full power *S* consumption. Different buildings consumptions have been analyzed, and the relevance of the chosen modeling by fBm increments has been evaluated using limit theorems. The used methodology is appropriated for short-term prognosis for all considered objects except of the Theater and Textile Industry.

ACKNOWLEDGMENT

The present survey has been carried out as a result of the collaboration between Bulgarian and French teams working together during the joint project "Cluster analysis of aperiodic time series signals" financed by the Bilateral program PHC "RILA" projects 2017-2018 – Bulgaria – France.

REFERENCES

[1] O. Barndorff-Nielsen. (1978). Hyperbolic distributions and distributions on hyperbolae. *Scandinavian Journal of Statistics*, vol. 5, № 3, pp. 151–157. Available: http://www.jstor.org/stable/4615705?seq=1#page_scan_tab_contents
[2] E. Eberlein. *"Application of generalized hyperbolic Lévy motions to finance"*, in Lévy Processes – Theory and Application, Springer International Company AG, 2001, pp. 319-336. Available: https://link.springer.com/chapter/10.1007/978-1-4612-0197-7_14

[3] O. E. Barndorff-Nielsen, N. Shephard. (2001).–" *Modelling by Levy processes for financial econometrics*", in Lévy Processes – Theory and Application, Springer International Company AG, 2001, pp. 283-318. Available:
https://link.springer.com/chapter/10.1007/978-1-4612-0197-7_13
[4] O. E. Barndorff-Nielsen, T. Mikosch. (2012). "*I. A tutorial on Levy processes*", "*II. Distribution, Pathwise and Structural Results*" in Lévy Processes – Theory and Application, Birkhäuser, Boston, 2001, pp. 1-106. Available:
https://books.google.bg/books?hl=bg&lr=&id=uInbBwAAQBAJ&oi=fnd&pg=PR7&dq=%5B4%5D%09O.+E.+Barndorff-Nielsen,+T.+Mikosch.+(2012).+L%C3%A9vy+processes:+theory+and+applications&ots=EaHSsdkb9-&sig=pZFIa799TAffeUyWDQRAFqNHhbs&redir_esc=y#v=onepage&q&f=false
[5] I. Nourdin, F. G. Viens. (2009). Density formula and concentration inequalities with Malliavin calculus, *Electronic Journal of Probability*, № 14, pp. 2287-2300.
[6] I. Nourdin, D. Nualart, C. Tudor. (2010). Central and non-central limit theorems for weighted power variations of fractional Brownian motion. *Annual of Institute H. Poincaré Probability Statistics*, vol. 46, № 4, pp. 1055-1079.
[7] Individual household electric power consumption https://archive.ics.uci.edu/ml/datasets.html?format=&task=&att=&area=&numAtt=&numIns=&type=ts&sort=nameUp&view=table

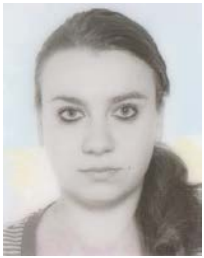

**Valeria Bondarenko** was born in Kiev, Ukrainian, in 1986. She received B.S., M.S. degrees and studied graduate school in Science in Applied Mathematics (System Analysis Control) from the National Technical University of Ukraine "KPI", Institute of Applied System Analysis. She defends a Ph.D. thesis in Engineering Science with the National Technical University of Ukraine "KPI", Institute of Applied System Analysis at 2017.

From 2006 to 2007 she works as a Research assistant, with the National Technical University of Ukraine "KPI", Department of Applied Mathematics, National Academy of Sciences of Ukraine. During the period 2007-2010 she is an Economist-Analyst at the Joint Stock Company the State Export-Import Bank of Ukraine (JSC Ukreximbank). After that from 2010 to 2014 she become a Doctoral assistant with the National Technical University of Ukraine "KPI", Institute of Applied System Analysis, National Academy of Sciences of Ukraine. During December 2014 – August 2016 she is a Research assistant with the National Technical University of Ukraine "KPI", Research Institute of Information Processes, National Academy of Sciences of Ukraine. From September of 2016 to June 2017 she makes a mobility with in the Institut de recherche en Communications et Cybernétique de Nantes IRCCyN, Ecole Centrale de Nantes. From October 2017 till now she is a Research assistant with in the University of Beira, Covilha. Her research interests include Econometrics and Time Series Analysis; Control theory; Modeling of Financial Securities Dynamics by means Diffusion Models (Various Diffusion Components of the Model); Applied Statistics; Labor Economics; Differential equations.

Mrs. Valeria Bondarenko awards include Cisco Certificate (2007), scholarship "Studienstiftung des Abgeordnetenhauses von Berlin" (2013-2014), Research intership, Lameta, Montpellier (2016).

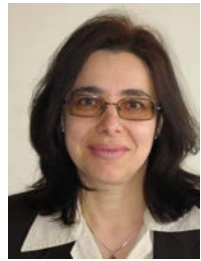

**Simona Filipova-Petrakieva** was born in Sofia, Bulgaria, in 1971. She finished Secondary School of Mathematics in Sofia, Bulgaria (1989) with the gold medal and she was on the top of graduated. She received B.S. and M.S. in Electronics and Automatics from the Technical University of Sofia, Bulgaria. She defended a Ph.D. thesis in Electrical Engineering, Electronics and Automatics, Theory of Electrical Engineering with the Technical University of Sofia, Bulgaria.

From 1994 to 2000 she works as an Assistant Professor with the department Industrial Automation, Faculty of Automation, Technical University of Sofia, Bulgaria. From 1994 to 2000 she also is an Editor-in-Chief of the journal PC Magazine – Bulgaria. From 2000 to 2009 she works as an Assistant Professor with the department Theory of Electrical Engineering, Faculty of Automation, Technical University of Sofia, Bulgaria. From 2009 till now she teaches there as an Associate Professor. Her research interests include Theory of Electrical Engineering, Neural Networks, Cluster Analysis, Graphs Theory, Interval Methods for Analysis and Synthesis of Linear Circuits and Systems, Discrete Event Systems, Discrete Structures in Mathematics, Chaotic Theory for Information Transmission, Brownian Motion Theory. Hers teaching activities become in courses: "Theory of Electrical Engineering" (2001 till now) and "Discrete structures in Mathematics" (2008 till now), Technical University of Sofia, Sofia, Bulgaria.

Mrs. Simona Petrakieva has excellent communication skills gain through the participation in ERASMUS+ teaching mobility program. She is manager and scientific advisor of national and international research projects, and reviewer for *Journal of Abstract and Applied Analysis, Journal of Applied Numerical Mathematics, Journal of Mathematical Communication, Arabian Journal for Science and Engineering, etc.*

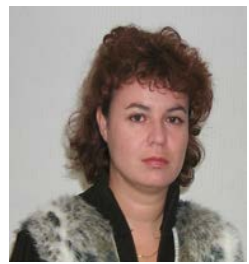

**Ina Taralova** received the Bachelor's degree in control engineering and the Master's degree in robotics, both from Sofia Technical University, France, and the Ph.D. degree in nonlinear system dynamics from INSA (Institut National des Sciences Appliquées), Toulouse, France, in 1996.

She became a Marie-Curie Post-Doctoral Fellow, in the Department of Electrical and Electronic Engineering, University College Dublin, Dublin, Ireland, in 1997, conducting research on nonlinear circuit dynamics. She then worked as a Research Associate at the Control Systems Centre, UMIST, Manchester, U.K.

Since 2000, she has been Assistant, and after Associate Professor, at Ecole Centrale de Nantes, Nantes, France, teaching control engineering, optimization, research methodology. She is currently a Researcher in the Laboratoire des Sciences du Numérique de Nantes (LS2N), which is a Joint Research Unit (U.M.R. 6004). Her scientific interests include control engineering, nonlinear system dynamics, pseudo-random chaotic generators design, secure information transmission. She is manager and scientific advisor of national and international research projects, and reviewer for *International Journal of Bifurcation and Chaos, Control System Letters, Control Engineering Practice; IMA Journal of Mathematical Control and Information etc.*

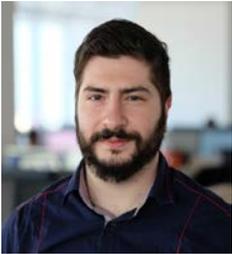

**Desislav Andreev** was born in 1991 in Dobrich, Bulgaria. He received his B.S. in Computer Systems and Technologies at the Faculty of Computer Systems and Control of Technical University of Sofia. Next he received his second M.S. in Computer Systems and Technologies at the Faculty of German Engineering and Economics Education of Technical University of Sofia in 2016. He was top of the graduated.

From 2012 to 2014 he works as a Software Developer in Atia Soft Ltd. and in the meantime starts as an Assistant Professor in the Faculty of Computer Systems and Control. I parallel he developed software for video editing tools, data monitoring and media database administration. From 2014 till now he work in Visteon Electronics as a Software Engineer. His main area of expertise is the system and infotainment states of automotive ECUs and he is one of the co-founders of the Visteon Engineering Academy. He engineered the architecture of the low-level software, communication layers, middle layer software and user interface protocols, thus assuming in 2016 the position of Senior Software Engineer. In 2016 he started as a Ph.D. student in Artificial Intelligence Systems at the Technical University of Sofia, preparing dissertation about Machine Learning Algorithms in Quantum Entanglement-based Environment. In 2017 he transferred his position in the ADAS department of Visteon, working on Autonomous Driving Technologies and becoming in 2018 lead of the Sofia branch of the department. He still works as a Assistant Professor. His main research interests are in Parallel Computing, Pattern Recognition, Sensor Fusion, Quantum Physics and Quantum Field Theory, Artificial Environments and World Simulations.

Mr. Desislav Andreev has excellent communication and leading skills gained through his work in Visteon as a lecturer, team lead and a software engineer. His job requires multiple business travels to Santa Clara, Detroit and Karlsruhe, working together with clients and partners of Visteon. He is an advanced developer in C and C++ programing languages.